\documentclass[12pt]{article}
\usepackage{amsmath,amssymb,graphicx,mathrsfs}
\newcommand{\be}{\begin{equation}}
\newcommand{\ee}{\end{equation}}
\newcommand{\bea}{\begin{eqnarray}}
\newcommand{\eea}{\end{eqnarray}}

\def\({\left(} \def\){\right)}

\begin{document}
\title{\vspace{-1.8in}
{Constraints on the quantum state of pairs produced by semiclassical black holes}}
\author{\large Ram Brustein${}^{(1)}$,  A.J.M. Medved${}^{(2,3)}$ \\
\vspace{-.5in} \hspace{-1.5in} \vbox{
 \begin{flushleft}
  $^{\textrm{\normalsize
(1)\ Department of Physics, Ben-Gurion University,
    Beer-Sheva 84105, Israel}}$
$^{\textrm{\normalsize (2)  Department of Physics \& Electronics, Rhodes University,
  Grahamstown 6140, South Africa }}$
$^{\textrm{\normalsize (3) National Institute for Theoretical Physics (NITheP), Western Cape, South Africa}}$
 \\ \small \hspace{1.7in}
    ramyb@bgu.ac.il,\  j.medved@ru.ac.za
\end{flushleft}
}}
\date{}
\maketitle
\begin{abstract}

The pair-production process for a black hole (BH) is discussed within the framework of a recently proposed  semiclassical model of BH evaporation.  Our emphasis is on how the requirements of unitary evolution and strong subadditivity act to constrain the state of the produced pairs and their entanglement with the already emitted BH radiation.  We find that the state of the produced pairs is indeed strongly constrained but that  the semiclassical model is consistent with all requirements.  We are led to the following picture:  Initially, the pairs are produced in a state of  nearly maximal entanglement amongst the partners, with a parametrically small entanglement between each positive-energy partner and the outgoing radiation, similar to Hawking's model. But, as the BH evaporation progresses past the Page time,  each positive-energy partner has a stronger entanglement with the outgoing radiation and, consequently, is less strongly entangled with its negative-energy partner. We present some evidence that this pattern of entanglement does not require non-local interactions, only  EPR-like non-local correlations.
\end{abstract}
\newpage

\section{Introduction}

In a recent series of papers \cite{slowleak,slowburn,flameoff,noburn,endgame,future,stick}, we have developed a new theory of semiclassical black hole (BH) evaporation, extending Hawking's seminal works \cite{Hawk,info} to finite-mass BHs. As motivated in
\cite{RJ,RM,RB,flucyou},  we have included the fluctuations of the background geometry by way of a wavefunction for the horizon of the incipient BH. Our guiding principle was to closely follow Hawking's original  calculations ---  but, instead of assuming  a  classically fixed (Schwarzschild) metric, we  evaluated background-dependent  quantities as quantum expectation values in the state
of the  BH. This prescription guaranties that, on average, the emission rate from a BH is the standard thermal rate and that the fluctuations about the average rate are small.   We have also included the classical time dependence of the particle emissions and their classical back-reaction on the BH.

The quantum corrections that arise from the background fluctuations are proportional to the inverse of the BH entropy, $1/S_{BH}$. This is contrary to the standard methods of  effective field theory in a fixed curved-space background, for which  one only expects exponentially small corrections from the quantum back-reaction of the matter fields on the classical geometry. But our framework goes beyond that of an effective field-theory description of BH evaporation. From the standard point of view, the corrections arising from  our model  would correspond to  non-perturbative contributions to expectation values that vanish to all orders in perturbation theory.

Let us  emphasize that BHs do not ``burn'' like normal black bodies do; rather, their burning (evaporation) is of a highly quantum nature. For example, the Sun and a BH at the same temperature will emit differently. The Sun  emits radiation in a highly classical state with high occupation numbers for the emitted modes,
whereas the BH, on the other hand, will emit a highly quantum state with low occupation numbers. (See, for instance, \cite{future}.) As Hawking showed in \cite{info}, these quantum emissions can (but do not have to) be described  in terms of pair production. Indeed, by adapting Schwinger's pair-production formula to the case of a BH, one obtains the correct rate of  emission (see below).

In \cite{flameoff}, we have described  BH evaporation  in our semiclassical framework from the  perspective of pair production.
 Just as in  Hawking's model, the negative-energy pair partners are  subsumed into the BH interior at a rate that is determined by the thermal rate of emission. But, for a BH of finite mass  --- and  contrary to the situation in Hawking's model of an  effectively eternal BH ---  this process acts to bound  the number of entangled pairs in the near-horizon zone at any given time, such that their number is parametrically smaller than the BH entropy. This is the key fact that allows information to  be released from the BH without over-exciting the state of the near-horizon region. However, in \cite{flameoff}, we assumed that the pairs are produced in a pure state or, equivalently, that the  produced pairs and the outgoing radiation are not entangled. Here, this assumption is relaxed and we in fact  show that the pairs must be entangled to some degree with the outgoing radiation.

Let us consider a single pair, which is produced in a process that is akin to a gravitational version of the  Schwinger effect.  Schwinger's famous equation \cite{Schwing}  predicts the rate per unit volume $R_{PP}^{\rm E}$ of electron--positron pair production in an electric field ${\cal E}$, where   $\;R^{\rm E}_{PP}= \frac{\alpha^2}{\pi^2 }{\cal E}^2  e^{-\frac{\pi m^2}{e {\cal E}}}\;$. Now suppose that  one substitutes the gravitational force $F_{\rm G}$ for the electric force $\;F_{\rm E}={\cal E}e\;$ in Schwinger's equation. If  $m$ denotes the relativistic mass of the positive-energy partner, the Newtonian  gravitational force is given by
$\;F_{\rm G}=\frac{G_NM_{BH}m}{r^2}$.~\footnote{Here, $G_N$ is Newton's constant, $\;M_{BH}=\frac{R_S}{2G_N}\;$ is the BH mass, $R_S$ is the Schwarzschild radius, and the speed of light and Boltzmann's constant are set to unity.} The resulting expression for the rate of  gravitational pair production per unit volume ${\cal R}^{\rm G}_{PP}$ in the near-horizon limit  is then
\be
{\cal R}_{PP}^{\rm G}\;\sim\;  \frac{\hbar}{R_S^4}\left(\frac{R_S m}{2\pi\hbar}\right)^2  e^{-\frac{2\pi R_S m}{\hbar}} \;.
\label{heur}
\ee

This rate is maximized when $\;m = \frac{\hbar}{\pi R_S}\;$, which is of the order of the Hawking temperature $\;T_H = \frac{\hbar}{4\pi R_S}\;$; in which case, $\;{\cal R}^{\rm G}_{PP}\sim\frac{\hbar}{R_S^4}\;$. Meaning that, as expected, one Hawking pair is produced  per light-crossing time $R_S$ from a volume $R_S^3$. Away from the horizon ($r>R_S$), the rate of pair production  becomes exponentially small, and so the idea that the quantum emission process originates from
pair production near the horizon is indeed supported.

Schwinger's formula and its gravitational analogue provide the rate for the pair-production mechanism but contain no information about the state of the produced pairs. In Hawking's model, the entanglement is maximal, corresponding to the original vacuum state in the vicinity of the horizon. In the absence of additional physics introducing new scales, this follows from the adiabaticity of the collapse process \cite{Hawk,PolchinskiJer}.  However, in our model, the shell (incipient BH) is fluctuating and so  adiabatic considerations need to be revised.  This issue was discussed for harmonic oscillators in \cite{flucyou}, but  we would still like to understand in detail how the situation is changed for our  semiclassical model. Another example in which adiabaticity is modified is given by the emission from fuzzballs \cite{SamirJmart}.

The overall goal of the current paper is to better understand the nature of the quantum state of the pairs that are produced in this gravitational Schwinger process in the context of our semiclassical model for  BH evaporation.

The dependence of the degree of entanglement of  pairs on the state of the producing field was addressed for  electron--positron pair production by an electromagnetic field in, for example,  \cite{fedorov,krekora}.  The state of a  pair can vary between maximally entangled and a product state, depending on the properties of the background field. In our case, we do not have such a detailed knowledge of the dependence of the state of the pairs on the production conditions. We will, instead, choose to parametrize the possible states and look for general constraints from the conditions of unitarity evolution and strong subadditivity (SSA) of entropy.

In the context of BH pair production, entanglement is expressed in terms of quantum correlations of the times,  frequencies and  (possibly) the polarizations of the emissions. Because we consider a Schwarzschild BH,  the angular momenta of the emitted particles has to sum up to zero and each particle has to be emitted within the thermal-frequency window. Consequently, the main variable whose quantum correlations determine the amount of entanglement is the emission time of the particles.  These correlations  determine the off-diagonal elements of both the single-particle density matrix for the emitted radiation and the density matrix for the produced pairs. In many discussions on the state of the emitted radiation, the entanglement is modeled in terms of spin degrees of freedom (qubits). One can also map the emission times formally into qubit states; however, it is important to remember that the true physical variables are different.

But  an outside observer can only determine the state of the outgoing radiation. She might want  to use her knowledge of this state  to determine the state of the produced pairs. However, it is clear that such a determination can only be achieved by supplying additional ingredients about the physics of the pair-production process. Here, we will not attempt to construct such a physical model but, rather, use quantum-information concepts to constrain the possible  models and to show that our semi-classical framework is indeed consistent with all the requirements.

The plan of the paper is as follows: The next section recalls some of
the  basic elements
 of our semiclassical model, as  needed for the rest of the discussion.  Then, in Section~3, we consider the constraints that  quantum theory imposes on the process
of BH evaporation,  on both general grounds and in the context
of our semiclassical picture.
In particular, we address  Mathur's argument about the conflict between unitarity and SSA \cite{Mathur1,Mathur2,Mathur3,Mathur}.  These ideas are
put on a quantitative level in Section~4, where the conditions of unitarity and SSA
are used to constrain the state of the pairs in our framework.
We are able to demonstrate that our model is consistent with both of these principles. The paper ends in Section~5 with a brief summary.

\section{Review of semiclassical finite-mass black holes}

Before proceeding, we will review some basic elements of  our semiclassical framework. A  four-dimensional Schwarzschild BH with radius $R_S$ and mass $M_{BH}$ is assumed for concreteness. All fundamental constants
are  set to unity except
for $\hbar$ and Newton's constant $G$, or equivalently,  the Planck length $l_P=\sqrt{\hbar G}$.

\subsection{The basics}

\begin{itemize}
\item{The classicality parameter $C_{BH}$}

We define $C_{BH}$ as the ratio of the Compton wavelength of the BH $\frac{\hbar}{2\pi M_{BH}}$ to its Schwarzschild radius $ R_S$, then $\;C_{BH}= \frac{l_P^2}{\pi R^2_S}=S^{-1}_{BH}\;.$ This parameter characterizes the deviation from a
classically fixed, curved spacetime ({\em i.e.}, $\;G\to 0\;$, $\;M_{BH}\to\infty\;$, $\;R_S\gg 1\;$ but finite), so that  the semiclassical regime is when $\;C_{BH}\ll 1\;$ but finite. The semiclassical corrections to physical quantities  typically come as a power series in $C_{BH}$. These corrections are non-perturbative from the point of view of an effective field theory in a fixed, curved background; hence, their inclusion is the essential difference between our analysis and the standard discussions in the literature.

  The  parameter $\;C_{BH}=C_{BH}(R_S(t))\;$ can be thought of  as a dimensionless, time-dependent $\hbar$. It is formally introduced into our theory as the dimensionless width (squared) of the BH wavefunction. The same parameter has appeared in a different guise in \cite{Dvali1,Dvali2} and  also corresponds to the (small) parameter $1/N^2$ in the AdS/CFT correspondence \cite{adscft} when $N$ is large but finite.

\item{The emission rate of Hawking radiation}

The Stefan--Boltzmann law determines the approximate classical time dependence of the Schwarzschild radius,  $\;R_S(t)=R_S(0)\left[1-\left(\frac{t}{\tau_{BH}}\right)^{1/3}\right]\;$, where $\tau_{BH}\simeq S_{BH}(0)R_S(0)\;$ is the BH lifetime.
We may now use $\;S_{BH}\sim R_S^2\;$ along with the relation between the number of particle emissions~\footnote{This use of $N$ should not be confused with the rank of the field theory
in AdS/CFT.} $N$
and the entropy,
$\;\Delta N=-\frac{\Delta M_{BH}}{E}\simeq -\frac{\hbar\Delta M_{BH}}{T_H}= -\Delta S_{BH}\;$, to obtain
$\;N(t)= S_{BH}(0)\left(\frac{t}{\tau_{BH}}\right)^{2/3}\;$.
This allows us to replace  $t$ by $N$ and use the latter
 as our (dimensionless) time coordinate. For instance,
the Page time \cite{page} is simply the ``time'' $N$ when
 $\;S_{BH}(N)=N\;$. The average emission rate of the semiclassical model is equal to the Hawking emission rate with small fluctuations, of order $C_{BH}$, about the average rate.

\item{The semiclassical single-particle density matrix}

The single-particle density matrix of the outgoing Hawking radiation
$\rho_{SC}$~\footnote{We previously called this matrix the  ``density matrix'' or ``radiation matrix".}  is the semiclassical two-point function or number operator, but with an appropriate normalization. For a free theory, the  full density matrix is completely determined by the corresponding single-particle density matrix. This is  already evident in Hawking's work \cite{info} and is further clarified in \cite{future}.

Each entry in   the single-particle density matrix depends in principle on the frequency, polarization and emission time of the radiated particles. We ignore the polarization dependence.  In the semiclassical model, the matrix
$\rho_{SC}$  no longer has Hawking's diagonal form  and, as a result, the evaporation process becomes unitary even though the thermal-like emission spectrum is kept \cite{endgame,future}. The matrix  $\rho_{SC}$ picks up off-diagonal contributions that are uniform in terms of frequency but suppressed relative to the diagonal elements by $C_{BH}^{1/2}(N)$ \cite{slowleak}.

The elements of $\rho_{SC}$ do have a non-uniform suppression in terms of emission
time; modes emitted at different times tend to decohere \cite{slowburn}. Nonetheless, if the radiation
is being regularly monitored at intervals of $\;\Delta N\sim \sqrt{S_{BH}}\;$ or less, then this suppression can be compensated \cite{endgame} (and see below).
Thus,  off-diagonal elements of the  matrix $\rho_{SC}$ can be regarded as uniform in magnitude with respect to both frequency and emission time.  Knowing this form allows us to calculate the R\'enyi entropy of the full density matrix \cite{future}.

As implicit in our earlier works and   explained in \cite{future}, $\rho_{SC}$ can be viewed as an $N\times N$ matrix, with the indices running over the wave-packet modes with non-vanishing occupation number and with the diagonal elements given by the average occupation number for each mode. The elements  can then be
expressed  to good
approximation   as \cite{slowleak,slowburn}
\begin{align}\label{rhosc}
&(\rho_{SC})_{ii}\; =\;  1 \; , \notag \\
&(\rho_{SC})_{i\neq j} \;=\; \sqrt{C_\text{BH}(N)} e^{i \theta_{ij}} \; ,
\end{align}
where  the phases $\theta_{ij}$ can be treated as random for most purposes.

\item{Tracking}

An observer is said to be  tracking the radiation  when  she monitors and records the amplitudes and phases of all the non-vanishing elements of  the
single-particle  density matrix $\rho_{SC}$ at regular time intervals. These time intervals should be short enough (as described above) to allow the observer eventually to  record all the entries of $\rho_{SC}$ and,
thus, reconstruct the full density matrix.  Then, via the off-diagonal elements, she will possess  knowledge about all the correlations of the emitted particles, even those that have since decohered.  She will also have to compensate for the classical time-dependence of $C_{BH}$ as explained in \cite{endgame}. It will always be assumed that the radiation is being tracked.

\end{itemize}

\section{Constraints from unitarity and strong subadditivity}

We will now address the apparent conflict between unitary evolution and SSA,  as has  often been emphasized by Mathur, and then, in the next section, explicitly verify that our model can satisfy  both of these conditions. Our discussion focuses on parametric dependence; hence,  numerical factors are sometimes
omitted for clarity.

\subsection{Unitarity}

It has long been accepted --- at least since the advent of the gauge--gravity duality  --- that an evaporating BH respects unitarity, as would be the case for any other process of  quantum-mechanical evolution.  The benchmark model for describing BH evaporation as a unitary process  is the Page model
\cite{page}. (Also see \cite{HaydenPreskill}.)
Page assumes that the BH starts off in a pure state and, at later times,
 views the remaining  BH as the ``purifier'' of the emitted radiation
in some random basis. Meaning that  a random unitary transformation
relates this basis to that in which the density matrix of
the composite BH--radiation system has a single entry.
With these assumptions, Page is  describing the minimal requirement
for a BH to release all of its information before
the end of evaporation.

\subsection{Strong subadditivity}

Here, we recall Mathur's arguments  \cite{Mathur1,Mathur2,Mathur3,Mathur}  about the SSA inequality and explain their implication to our framework.
Mathur's perspective is  closely related to but distinct from
that of the ``firewall'' proponents \cite{AMPS}  (also, \cite{Sunny1,Braun,Bousso2}). In brief,  Mathur assumes a  unitary model of BH evaporation which allows for only small corrections to Hawking's picture and  contends that, for the system consisting  of entangled pairs plus escaped Hawking particles, the entanglement entropy will grow monotonically  throughout the pair-production process.

Mathur starts by  assuming that the pairs are produced in a state which is approximately pure and that the pair partners are approximately in a state of  maximal
entanglement. This implies that the pairs are produced in a state which is approximately equal to that of the Hawking model. The deviations of the pair state from maximal entanglement is parametrized by a small parameter $\epsilon$. Specifically, each of the produced pairs is assumed to have an associated  entanglement entropy in bits of $\;S_{pair}=1-\epsilon\;$, where $\epsilon$ is meant as a small number. Mathur then applies the SSA inequality  to conclude  that, after the production of  $N$ pairs, the total entropy of the outgoing radiation is bounded from below  by $NS_{pair}$. And so, as long as deviations from Hawking's model are small ({\em i.e.}, $\;\epsilon\ll 1$), the entanglement entropy is necessarily large. Mathur correctly observes that a monotonically growing  entanglement entropy is contrary to the behavior of a ``normal'' burning body.

The key assumption which is made by Mathur is that $\epsilon$ is roughly constant and small for each of the produced  pairs. In particular, $\epsilon$
is assumed to  have at most a weak dependence on the history of the BH;  meaning that it is essentially an $N$-independent number. The constancy of $\epsilon$ is attributed to locality. It will be shown that this critical assumption is modified in our framework and indications will be given that non-local interactions are {\em not} needed, only non-local correlations.

Let us next recall the precise meaning of SSA; first in general and  then in the current context. The SSA inequality relies on the unitarity of quantum mechanics and is a statement about  a tripartite quantum system $|A\rangle\otimes|B\rangle\otimes|C\rangle$. It asserts that the
associated von Neumann  entropies
must satisfy the bound \cite{SSA}
 $\; S_{AB} + S_{BC} \geq S_{ABC} + S_{B}  \;$ or, equivalently,
\be
S_{AB} + S_{BC} \;\geq\; S_{A} + S_{C}\;.
\ee
Here, $\;S_X = -{\rm Tr}_{X}[\widehat{\rho}_X\ln{\widehat{\rho}_X}]\;$ such that $\widehat{\rho}_X$ is the reduced density matrix for subsystem
$X$. For instance,
 $\;\widehat{\rho}_{AB}={\rm Tr}_{C}[\widehat{\rho}_{ABC}]\;$, $\;\widehat{\rho}_{A}={\rm Tr}_{BC}[\widehat{\rho}_{ABC}]\;$ and so forth. Equality is obtained if
and only if  $ABC$ is in a pure state,  $\;S_{ABC}=0\;$.

Now, in the context of BH pair production, Mathur takes subsystem $A$ to be the positive-energy modes that have already moved far from
the BH and $B$ and $C$ to be, respectively, the positive- and negative-energy modes of a newly formed entangled pair (see Fig.~1).
If  $\epsilon$ is indeed approximately constant and small, it is clear that $\;S_{BC}=0\;$ up to irrelevant corrections and  the reduced density matrix is
then  ``thermal",  $\;S_{B}=S_{C}=1\;$. Hence,
the bound reduces to
\be
S_{AB}  \geq S_{A} +1  \;.
\ee

\begin{figure}
[t]
\begin{center}
\scalebox{.4} {\includegraphics{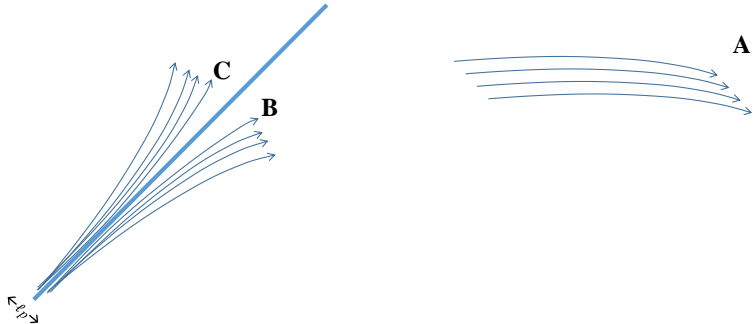}}
\end{center}
\caption{The $ABC$ system shown on a spacetime diagram. The outgoing Hawking particle $A$ is far from the BH horizon while the $B$ and $C$ modes are near the horizon. }
\end{figure}

Now suppose that this newly formed pair is the $N^{{\rm th}}$  such pair produced. It would follow, according to Mathur,
 that $S_{A}=N-1$ (as each
escaped Hawking mode is still presumed to share nearly maximal entanglement with its partner) and thus
\be
S_{AB}\geq N \;.
\ee

Extrapolating this process to the Page time, when the BH entropy is reduced to half its original size \cite{page}, one obtains
\be
 S_{rad} \geq \frac{1}{2}S_{BH}(0)\;,
\ee
where $S_{rad}$ means the entropy of the external Hawking radiation. Mathur then contends that the process
continues in this fashion until the BH can no longer be regarded as semiclassical. But, by then, it is too late to make  $S_{rad}$ small
as the remaining BH lacks  the entropy storage capacity to purify the radiation.

Mathur has argued~\footnote{See, especially, the discussions about locality in \cite{Mathur2}.} that,  if one restricts the discussion to the context of effective field theory in a fixed background, then the small parameter $\epsilon$ cannot depend on the number of emitted particles $N$ without an accompanying violation of locality. This is because the effects of a strictly local interaction  can not depend on the characteristics nor the number of the  previously  emitted Hawking modes. We believe that this argument is indeed correct within the realm of  effective field theory on a fixed, curved background. However, we will argue --- pending further investigations --- that, when the background is treated as a quantum state and its fluctuations are taken into account, such $N$-dependence is the result of EPR-like, non-local correlations rather than non-local interactions.

The essence of our argument is that  the degree of entanglement could depend on the state of the BH (and possibly  on the state of the pairs near the horizon), as typically happens in normal quantum systems. And, given unitary evolution,  the state of the BH must depend on the state of the outgoing radiation. In fact, an example of such a phenomenon was given by Chowdhury and Mathur in the context of a fuzzball model \cite{SamirJmart}. There, the early emissions of the Hawking particles indeed change the state of both the BH and the near-horizon matter such that later emissions are  correlated with the early ones.  In contrast, our semiclassical model, as of now, does not include the physics that allows us to estimate the degree of entanglement of the pairs. We then have to resort to general arguments and constraints.

It is standard, classically or semiclassically, to regard the entangled pairs as being created in a pure state.
Our claim, however, is  that the pairs are produced in an entangled state with the outgoing radiation and so cannot, by themselves, form a pure state. Initially, though, the entanglement of the pairs with the outgoing radiation is small and
the deviation of the pair partners from maximal entanglement can be estimated.
What  we find is that the value of $\epsilon$ is not necessarily constant. Indeed, our model suggests that, at early times much before the Page time,
\bea
\epsilon(N)\; &\lesssim& \;\frac{2 N C_{BH}}{(1+N C_{BH})}\;=\;\frac{2 N }{S_{BH}(0)}\
\label{epsN}
\eea
as follows from Bound~(\ref{epsiN}) below. Besides the explicit dependence on  $N$, one can also notice that $\epsilon$ can become
large (order one) when approaching the  Page time.

 As will be shown later, the pairs are indeed significantly entangled with the outgoing radiation after the Page time; in which case, the parameter $\epsilon$
will have lost its meaning.  This result suggests that, at times later than the Page time, the pairs are produced in a state that is becoming approximately a product state.

Now, to understand  the origin of  $\;\epsilon=\epsilon(N)\;$, let us
consider the perspective of a tracking observer.
Such an  observer does not have direct knowledge about the state of the pairs. She may believe that it is Hawking's state of maximal entanglement plus some small corrections, but this is only a guess ---  just like, in the standard EPR situation,  if Alice were to guess the spin of Bob's particle before even performing a measurement on her own. But,  because of the entanglement between the external particles and the pairs, a tracking observer will naturally acquire knowledge about the near-horizon state. The more measurements that she performs, the more precise her  knowledge about the state will become. At the same time, each such  measurement will project the state of  the pairs further away from a maximally entangled state.  If the observer is continuously monitoring the radiation, then the number of measurements must be of order $N^2$.

The fact that the number of emissions at the Page time is of order of the BH entropy has another important consequence. So many emissions can change the state of the BH itself, even though  each one on its own  makes only a small change.
This is  because the emissions are coherent and, therefore, their effects can accumulate. Consequently, the deviation of the state of the pairs near the Page time from the early-time maximally entangled state could also scale with $N$, without requiring non-local interactions. From the perspective of an  effective field theory in a fixed background, it would be impossible to see this effect at any order in perturbation theory. As such is the  case,  Mathur's argument about the necessity of non-local interactions would actually be correct.

One  might  still be tempted to attribute the $N$ dependence of $\epsilon$  to non-local interactions rather than entanglement. However, if this were so, then the modifications of  $\epsilon$ would be expected to depend on $N$ times  some coupling of the fields to the BH rather than be a function of $NC_{BH}$.

\section{Constraining the semiclassical model}

Of course, the  previous  discussion is a moot point if the evaporation process fails to be unitary or if the SSA inequality is invalidated. It is, therefore, worth checking that our semiclassical model is able to fulfill these minimal requirements.

A first test for any candidate model of unitary BH evaporation would be a demonstration that the rate of information release from the BH is at least fast as that predicted by the Page model \cite{Strominger}. Remarkably, our semiclassical model with tracking  passes this first  hurdle, as can  be observed in \cite{endgame,future}. We now want to understand this information-transfer process from the pair-production perspective.

Given that unitary evolution is indeed viable, we will use the SSA inequality to constrain the state of the pairs, as well as the entanglement pattern between the pairs and the outgoing Hawking particles.   Our model obeys the resulting  constraints, but  we still hope to be able to  derive the state of the pairs from a physical model of pair production in future research.

For the purposes of this presentation, we will follow  Mathur's argument and take subsystem~$A$ to be the escaped (early) Hawking particles, $B$ to be a positive-energy mode in the near-horizon zone (or late Hawking radiation) and $C$ to be its negative-energy partner in the zone. Although only  one of the pairs
need be considered, we  could just as well work with  a fraction of the pairs
and arrive at the same conclusions.

\subsection{Unitarity}

An essential point that we would like to reemphasize is that the pair-production picture is inherently ambiguous as far as an external observer is concerned \cite{info}. Such an observer can only  know for sure about   what she  learns by measuring the emitted particles; namely, the state of the outgoing  radiation.

This state was determined  in \cite{future}, where the density matrix of the external radiation $\widehat{\rho}$ was evaluated. This led to a calculation
of  the R\'enyi entropy  $\;H_2(\widehat{\rho})\;$, which is the main quantity that will be used to constrain the state of the pairs.  This entropy is, up to  subdominant corrections,
\be
H_2(N)\; =\; \frac{N}{1+N C_{BH}}\;=\;\frac{N(S_{BH}(0)-N)}{S_{BH}(0)}\;.
\label{renyiA}
\ee
It is  worth mentioning that these expressions for the R\'enyi entropy have higher-order corrections in $C_{BH}$ {\em but not} in $NC_{BH}$. Consequently,  any of our findings are  accurate as long as $C_{BH}=1/S_{BH}\ll 1$, which is assured until the  final stages of evaporation.

The R\'enyi entropy stops growing  when  $\;NC_{BH}=1\;$ ({\em i.e.}, at the Page time)
and then decreases  monotonically  for the remainder of the BH's lifetime.
That the  entropy depends  on $N$ in just this way  is indicative of a unitary process of evaporation, but it is also the main aspect of  Mathur's argument that unitarity is in contradiction with the SSA constraint.

\subsection{Strong subadditivity}

To check the status of the SSA condition, we consider the bound
\be
 H_2(A) + H_2(C)\;\leq\; H_2(AB) + H_2(BC) \;,
\label{inequal1}
\ee
which also implies the bounds
\be
 |H_2(A) - H_2(B)|\; \leq\; H_2(AB)\; \leq \; H_2(A) + H_2(B)\;.
\label{inequal2}
\ee
Bound~(\ref{inequal1}) is saturated when the $ ABC $ system is in a pure state, and a sharp inequality is expected when $ ABC $ is mixed.  We may use the R\'enyi entropy, rather than the von Neumann entropy, because the
Hawking modes are in a Gaussian state \cite{info,future} and the  SSA inequality  is respected by the R\'enyi entropy for such a state \cite{SSA1} as a  consequence of the Hadamard--Fisher inequality \cite{SSA2}.

As a  measure of entanglement, we will use  $E(X|Y)$, the (negative of the) conditional entropy \cite{Plenio}.  This is actually a lower bound on the true entanglement between $X$ and $Y$. The conditional R\'enyi entropy of the $AB$ system is given by
\be
E(B|A)\; =\; H_2(A)-H_2(AB)\;
\label{eba}
\ee
and, similarly, for  the $BC$ system,
\be
E(B|C)\; =\; H_2(C)-H_2(BC)\;.
\ee

The SSA inequality~(\ref{inequal1}) tells us that
\be
E(B|A)+E(B|C)\; \le\; 0\;,
\label{monogamy}
\ee
which is essentially a statement of ``monogamy of entanglement''.
For instance, if $B$ is strongly entangled with $C$, then $\;E(B|C)\sim 1\;$ and so  $\;E(B|A)\lesssim -1\;$,  implying that  $B$ is  weakly entangled with $A$
(and {\em vice versa}).

We  will first be considering  the constraints at times later than the Page time  when $\;N C_{BH}=1\;$.
A discussion about earlier times  then follows.

We begin here with a relation that follows from Eqs.~(\ref{renyiA}) and~(\ref{eba}),
\bea
E(B|A)&=&H_2(N)-H_2(N+1) \nonumber \\
&=& \frac{2N+1}{S_{BH}(0)}-1\;\simeq\;\frac{2N}{S_{BH}(0)}-1
\nonumber \\
&=& \frac{N C_{BH}-1}{1+N C_{BH}}\;.
\label{EBA}
\eea
Since $H_2(N)$ starts to decrease only after the Page time,  it is only after
this time that  $E(B|A)$ becomes positive and the SSA inequality becomes useful.  The measure of  entanglement $E(B|A)$ then grows monotonically from zero at $\;N C_{BH}= 1\;$ to  unity when $\;N C_{BH}\gg 1\;$. This tells us that the positive-energy partner $B$ is becoming more strongly entangled with the outgoing radiation as the evaporation proceeds.

From Bound~(\ref{monogamy}), we also find that
\be
E(B|C) \;\leq\; - E(B|A) \;=\; \frac{1-N C_{BH}}{1+N C_{BH}}\;.
\label{inequal3}
\ee
After the Page time,  the right-hand side of this bound is smaller than zero
and, therefore,
the conditional  entropy $E(B|C)$ is negative.
The entanglement between $B$ and $C$  is then getting weaker after the Page time; in particular, the  $BC$ system has to deviate significantly from a pure and maximally entangled state. Eventually, for $\;N C_{BH}\gg 1\;$,
the difference $H_2(BC)-H_2(C)$ has to be of order 1, which means that the $BC$
``pair'' has to be substantially mixed ---  essentially, a product state as
then $\;H_2(BC)\sim H_2(B)+H_2(C) \sim 2\;$.
This stands to reason because, as seen above, the positive energy-partner  already has a strong entanglement  with $A$.

We can also use this formalism to place limits  on the ``small'' parameter
$\epsilon$ from the previous section.
Since $\;H_2(A)\geq H_2(B)\sim 1\;$, at least until the late stages of the evaporation, it follows from
the left-most relation in Bound~(\ref{inequal2}) that
\be
H_2(B)\;\geq \; H_2(A)-H_2(AB)\;=\; \frac{N C_{BH}-1}{1+N C_{BH}}\;.
\ee
Then, parametrizing $\;H_2(B)=1- \epsilon (N)\;$, we find that
\be
\epsilon(N)\;\leq\; \frac{2}{1+N C_{BH}}\;=\; \frac{2 (S_{BH}(0)-N)}{S_{BH}(0)}\;.
\ee

As one can see, $\epsilon(N)$ is large (order unity) close to the Page time,
when $B$ can be expected to have significant entanglement with both
$A$ and $C$.
On the other hand, if $\;N C_{BH}\gg 1\;$, then $\;\epsilon(N)\ll 1\;$. This is another indication that the $AB$ system is becoming approximately pure and maximally entangled at times well past the Page time. Then, by  monogamy of
entanglement, the amount of
entanglement between the pair partners has to decrease accordingly.

Let us now discuss the situation before the Page time. At these times,
 the SSA inequality is not particularly  useful unless additional information
about the pairs is provided.
This is because of the minimal amount of entanglement between the pairs
and the external radiation. This is all that an external observer can truly know
about the state of the pairs at early times --- as far as she is concerned, the
$BC$ system could just as well be in a product state, a  maximally entangled state or  somewhere in between.

 However, we do know that, at early enough times, the semiclassical model can
be viewed as Hawking's plus perturbatively small corrections.
Hence, it can still be expected that the $BC$ system is approximately pure
and maximally entangled  up to corrections of order $\;N C_{BH}\ll 1\;$.

And, in spite of the  built-in ambiguity,  we can still address the size of
$\epsilon$ at early times.
The right-most relation  in Bound~(\ref{inequal2}) leads to
\be
H_2(B)\;\geq\; \frac{1-N C_{BH}}{1+N C_{BH}}\;.
\ee
Again parametrizing $\;H_2(B)=1- \epsilon (N)\;$, we then obtain
\be
\epsilon(N)\;\leq\; \frac{2 N C_{BH}}{1+N C_{BH}}\;=\;\frac{2 N}{S_{BH}(0)}\;.
\label{epsiN}
\ee
This fits very well with the expected behavior of the $BC$ system, being approximately pure and maximally entangled at early times and deviating from this
picture as the Page time is approached.

\subsection{A proposal for the entanglement}

From the previous discussion, one can observe the pivotal role that is  played by the Page time for which $\;NC_{BH}=1\;$. We expect that the amount of entanglement between the outgoing radiation and the produced pair is very small
at early times and large (order unity) at the later stages of evaporation.
 This suggests that, initially, the positive-energy horizon mode is almost in a product state with the outgoing radiation and almost maximally entangled with its negative-energy partner, $\;H_2(BC)\ll 1\;$.
 But, well after the Page time, when the horizon mode is significantly entangled with the outgoing radiation,
it is almost in a product state with its negative-energy partner,
$\;H_2(BC)\sim 1\;$. We further expect that, at intermediate stages close to the Page time, the horizon mode is
substantially entangled with both its partner and the outgoing radiation, but
with  the sum  of these two  entanglements bounded by unity from above.

We can make the above expectations more quantitative by adding the assumptions that the amount of entanglement is symmetric under the exchange $\;N \leftrightarrow S_{BH}(0)-N\;$ or, equivalently, $\;N C_{BH} \leftrightarrow \frac{1}{N C_{BH}}\;$ (see Section 5.3 of \cite{future}) and that
the total entanglement is fixed on the grounds of unitary evolution.
On this basis, we can propose that the following equalities are valid at {\em all
times} ({\em cf}, Eq.~(\ref{EBA})):
\be
E(B|C)\;=\; -E(B|A) \;=\;  \frac{N C_{BH}-1}{1+N C_{BH}}\;.
\ee
It is straightforward to check that all of the above expectations are
realized by  this proposal.

\section{Summary and Conclusion}

In brief, we have revisited the pair-production picture of
our semiclassical model of BH evaporation and have shown that
it can be constrained in a way that is  consistent with both unitary evolution and the SSA constraint on  entropy ({\em i.e.}, the  monogamy of entanglement).
The SSA inequality has proven to be a useful tool for constraining the state of the near-horizon modes,
but a physical description of the process is still lacking in our model. We hope to address
this matter at a future time.

Our findings are  in general agreement with the arguments of Mathur,
who argues that, within the realm of an effective  theory of fields on a fixed background geometry,  the  corrections to Hawking's model are small and cannot depend on the number of emitted particles  without violating locality.  However, when additional non-perturbative effects due to quantum fluctuations of the BH itself are taken into account, these assumptions are no longer in effect. Indeed,  the corrections from our model fail to satisfy  both of Mathur's conditions. However, we argue  that this violation  can be attributed to non-local entanglement rather than non-local interactions. Further analysis will be required to settle this matter conclusively.

The SSA inequality  necessitated an entanglement between
the positive-energy pair partners and the external Hawking radiation; the
degree of which becomes substantial after the Page time. It would then seem
that a (so-called) firewall would be an inevitable feature in  our model.
Nonetheless, another consequence of this framework
is an upper bound on the number of Hawking pairs in the near-horizon zone
that is parametrically smaller than the BH entropy. This  bound also limits the degree of excitement of the near-horizon state. The details of this
argument are
clarified elsewhere  \cite{noburn,stick}.

\section*{Acknowledgments}

We thank Doron Cohen, Samir Mathur and Daniel Rohrlich for useful discussions and Samir Mathur for enlightening comments on the manuscript. The research of RB was supported by the Israel Science Foundation grant no. 239/10. The research of AJMM received support from an  NRF Incentive Funding Grant 85353 and Competitive Programme Grant 93595, as well as Rhodes Discretionary Grants. AJMM thanks Ben Gurion University for their  hospitality during his visits.

\end{document}